\documentclass[preprint,12pt]{elsarticle}
\usepackage{amssymb}
\usepackage{float}
\usepackage{amsmath}
\usepackage{amssymb}

\usepackage{lineno,hyperref}
\modulolinenumbers[5]

\usepackage{xcolor}
\hypersetup{
    colorlinks,
    linkcolor={blue},
    citecolor={blue},
    urlcolor={blue}
}

\usepackage{lipsum}
\makeatletter
\def\ps@pprintTitle{%
 \let\@oddhead\@empty
 \let\@evenhead\@empty
 \def\@oddfoot{}%
 \let\@evenfoot\@oddfoot}
\makeatother

\begin{document}

\begin{frontmatter}

\title{Orbit Estimation Using a Horizon Detector in the Presence of Uncertain Celestial Body Rotation and Geometry}

\author{Amir Shakouri\fnref{label01}}
\fntext[label01]{Research Assistant, Department of Aerospace Engineering, \href{mailto:a_shakouri@ae.sharif.edu}{a\_shakouri@ae.sharif.edu}}
\author{Mahdi Hazrati Azad\fnref{label02}}
\fntext[label02]{PhD Student, Department of Aerospace Engineering, \href{mailto:hazrati_mhdi@ae.sharif.edu}{hazrati\_mhdi@ae.sharif.edu}}
\author{Nima Assadian\fnref{label03}}
\fntext[label03]{Associate Professor, Department of Aerospace Engineering, \href{mailto:assadian@sharif.edu}{assadian@sharif.edu}}
\address{Sharif University of Technology, 145888 Tehran, Iran}

\begin{abstract}

This paper presents an orbit estimation using non-simultaneous horizon detector measurements in the presence of uncertainties in the celestial body rotational velocity and its geometrical characteristics. The celestial body is modelled as a tri-axial ellipsoid with a three-dimensional force field. The non-simultaneous modelling provides the possibility to consider the time gap between horizon measurements. An unscented Kalman filter is used to estimate the spacecraft motion states and estimate the geometric characteristics as well as the rotational velocity of the celestial body. A Monte-Carlo simulation is implemented to verify the results. Simulations showed that using non-simultaneous horizon vector measurements, the spacecraft state errors converge to zero even in the presence of an uncertain geometry and rotational velocity of the celestial body.
\end{abstract}

\begin{keyword}
Orbit estimation \sep Horizon detector \sep Asteroid

\end{keyword}

\end{frontmatter}

%\linenumbers

\section{Introduction}
\label{S:1}

Autonomous orbit estimation is a key element of modern space missions. For planet Earth, the use of the Global Positioning System (GPS) for the orbital navigation at low altitudes \citep{Psiaki01,Beaudoin02,Sun03} is conventional. For high altitude missions the use of similar constellation-based navigation methods is proposed and tested as well \citep{Capuano04,Marmet05}. However, the use of GPS does not make the satellite completely autonomous, since it is related to the constellation of the GPS satellites and the constellation is mostly navigated from ground stations \citep{Cornara06}. On the other hand, relative states of two (or more) satellites can be utilized for an orbit estimation, independent of GPS satellites and/or ground stations \citep{Abusali07,Fakoor08,Psiaki09,Chang10,Tang11}. Additionally, natural properties of a planet, like its magnetic field \citep{Wu12,Farahanifar13}, atmosphere \citep{Ning14,Chen15}, or moons \citep{Stastny16}, can help to build an autonomous orbit estimation procedure. Spacecraft navigation and determination of Celestial Body (CB) characteristics can be autonomously accomplished using the planet’s geometric characteristics \citep{Lightsey17} or gravity field estimation \citep{Sun18}. 

Horizon detectors are known for their ability of determining the nadir vector. For nadir-pointing satellites, the nadir vector is frequently utilized as a measurement to estimate the attitude \citep{Serradeil19}. Furthermore, the nadir vector can be used to estimate the satellite orbit as well. For Earth orbiting satellites, horizon detectors have been used for orbit determination purpose assuming spherical \citep{Nagarajan20,Nagarajan21,Knoll22,Fitzgerald23,Hicks24} and non-spherical Earth models \citep{Li25}. Moreover, horizon sensors can be employed for finding the solar direction as discussed in \citep{Manassen26}. 

In this paper, an autonomous orbit estimation using discrete non simultaneous horizon detector measurements is addressed. Additionally, it is shown that these measurements can be utilized in the estimation of CB parameters; such as the semi-principal axes lengths and the angular velocity. The CB is modeled as a tri-axial ellipsoid, which is acceptable for most CBs in the solar system. The Unscented Kalman Filter (UKF) \citep{Ning27,Choi28,Kandepu29} is utilized for the estimation of the state and parameters in the presence of sensors noise and disturbances. The performance of this state and parameter estimation has been verified by the Monte-Carlo simulation. Thus, the main contributions of the paper are: (1) Unlike the previous investigations the time delays between horizon vector measurements are included, so the measurements are non-simultaneous; (2) the CB is modelled as a tri-axial ellipsoid with uncertain geometric characteristics that are augmented to the process model and estimated using parameter estimation; (3) similarly, the rotation of the CB about its primary axes is considered as an unknown and estimated in the filtering procedure; (4) MacCullaghs' formula \citep{Lowrie30} is assumed as the governing gravitational dynamic model in the three-dimensional force field; (5) for such a problem a measurement model is proposed as an algorithm and UKF is utilized to overcome the nonlinearities. 

The rest of the paper is organized as follows: First, the process model is formulated using relative dynamics and MacCullaghs' formula as the gravitational model. Next, the measurement model is derived and proposed as a unified algorithm. Section \ref{S:4} reviews the UKF algorithm and Section \ref{S:5} includes the simulation results. Finally, concluding remarks are presented.

\section{Process Model}
\label{S:2}

It is assumed that the geometry and the rotational velocity of the CB are not exactly known. Thus, by assuming the CB is a tri-axial ellipsoid, the semi-principal axes lengths ($a,b,c$), and its rotational velocity vector ($\boldsymbol\omega$) are included in the state vector of the system for the estimation purpose. In this manner, the process model can be summarized as the following equation:
\begin{equation}
\label{eq:1}
\dot{\boldsymbol x} = \boldsymbol f(\boldsymbol x)+\boldsymbol w
\end{equation}
where $\boldsymbol x=[\boldsymbol r^T \quad \dot{\boldsymbol r}^T \quad \boldsymbol \omega^T \quad a \quad b \quad c]^T$ is the state vector including $\boldsymbol r$ and $\dot{\boldsymbol r}$ as the position and velocity vectors of the spacecraft from the CB center of mass \citep{Scheeres31}. The state vector is augmented by the CB angular velocity and its semi-principal axes lengths to be estimated in the filtering procedure. A Gaussian, zero-mean white process noise, $\boldsymbol w$$\sim$$\mathcal{N}([0]_{12\times 1},Q)$, with a time-invariant covariance, $Q$, is linearly added to the system of equations. The vector function, $\boldsymbol f(\boldsymbol x)$, as the system differential equation is defined as
\begin{equation}
\label{eq:2}
\boldsymbol f(\boldsymbol x)=
\left (
  \begin{tabular}{c}
  $\dot{\boldsymbol r}$ \\
  $-\frac{\mu}{\|\boldsymbol r\|^3}+\boldsymbol a_{eul}+\boldsymbol a_{cor}+\boldsymbol a_{cen}+\boldsymbol a_{dis}$ \\
  $J^{-1}\boldsymbol \omega\times J\boldsymbol \omega$ \\
  $[0]_{3\times1}$
  \end{tabular}
\right )
\end{equation}
in which the state vector $\boldsymbol x$ is defined in a coordinate system associated with frame $A$, attached to the CB, a Celestial Body-fixed Coordinate System (CBCS). The Euler acceleration resulting from angular acceleration, $\boldsymbol a_{eul}=-\dot{\boldsymbol \omega}\times \boldsymbol r$ the Coriolis acceleration, $\boldsymbol a_{cor}=-2\boldsymbol \omega\times\dot{\boldsymbol r}$ and the centrifugal acceleration, $\boldsymbol a_{cen}=-\boldsymbol \omega\times(\boldsymbol \omega\times \boldsymbol r)$, are added to the two-body dynamics. The angular velocity $\boldsymbol \omega\equiv\boldsymbol \omega^{A/I}$ is defined as the rotation of the frame $A$ with respect to the inertial frame, $I$. The disturbance acceleration, $\boldsymbol a_{dis}$, is defined using MacCullaghs' formula \cite{Lowrie30}: 
\begin{equation}
\label{eq:3}
\boldsymbol a_{dis}=G\left(\frac{3}{2}\frac{\textrm{tr}(J)}{\|\boldsymbol r\|^5}\mathbb{I}_3+3\frac{J}{\|\boldsymbol r\|^5}-\frac{15}{2}\frac{\boldsymbol r^TJ\boldsymbol r}{\|\boldsymbol r\|^7}\mathbb{I}_3\right)\boldsymbol r
\end{equation}
for a CB with moments of inertia matrix $J$.

\section{Measurement Model}
\label{S:3}

\begin{figure}[H]
\centering\includegraphics[width=0.3\linewidth]{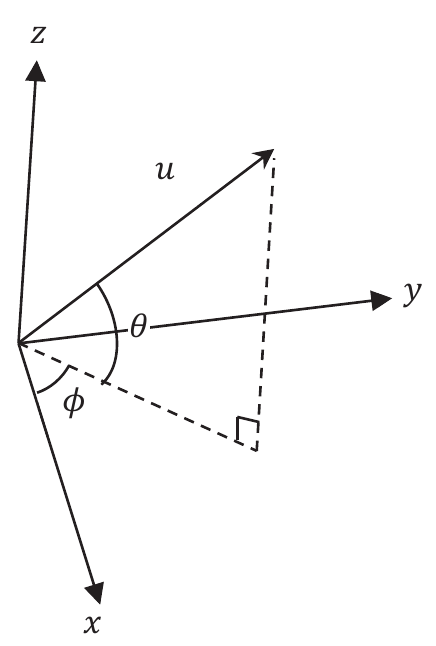}
\caption{Defining vector $\boldsymbol u$ in terms of spherical angles.}
\label{fig:1}
\end{figure}

The horizon sensor is used for the purpose of this study. Thus, the measurement is based on the horizon unit vector defined in an inertial coordinate system, $\boldsymbol u$. It is assumed that the attitude of the satellite has been determined by alternative sensors such as star trackers and is perfectly known. Thus, the horizon unit vector can be found in the inertial frame. This horizon unit vector is modeled by a pair of spherical angles. Therefore, the measurement model can be written as
\begin{equation}
\label{eq:4}
\boldsymbol z=\boldsymbol h(\boldsymbol x)+\boldsymbol v
\end{equation}
in which $\boldsymbol z=[\theta \quad \phi]^T$ is the measurement output vector, and is defined to be the spherical angles of $\boldsymbol u=[u_x \quad u_y \quad u_z]^T$:
\begin{equation}
\label{eq:5}
\boldsymbol u=
\left (
  \begin{tabular}{c}
  $\cos{\theta}$ $\cos{\phi}$ \\
  $\cos{\theta}$ $\sin{\phi}$ \\
  $\sin{\theta}$
  \end{tabular}
\right )
\end{equation}
as $\phi=$ tan$^{-1}(u_y/u_x)$ and $\theta=$ sin$^{-1}(u_z)$ (Fig. \ref{fig:1}). The measurement Gaussian zero-mean white noise in Eqn. \eqref{eq:4}, $\boldsymbol v\sim \mathcal{N}([0]_{2\times1},R)$, has a time-invariant covariance $R$. 

\begin{figure}[H]
\centering\includegraphics[width=0.5\linewidth]{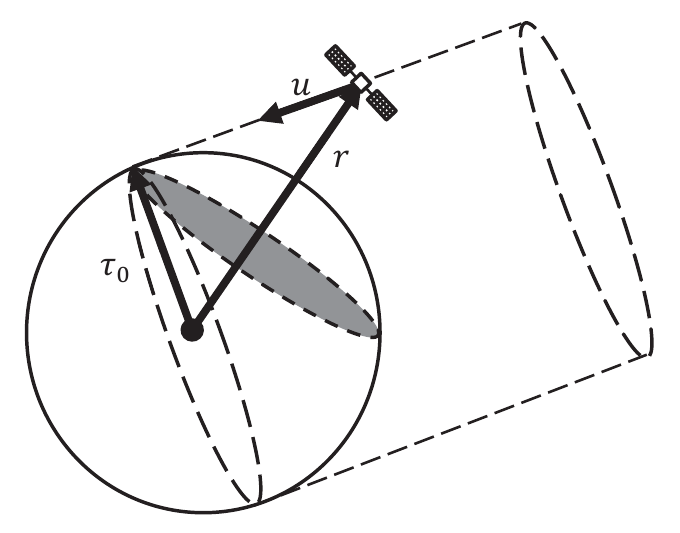}
\caption{A horizon unit vector measurement, $\boldsymbol u$, restricts the satellite position on an elliptic cylinder.}
\label{fig:2}
\end{figure}

In order to define the measurement model, $\boldsymbol h(\boldsymbol x)$, the formula of the horizon vector, $\boldsymbol u$, as a function of the position vector of the satellite, $\boldsymbol r$, should be found.  If the unit vector, $\boldsymbol u$, is measured from the satellite at the point, $\boldsymbol r$, toward the ellipsoid horizon, the satellite position should be located on a quadratic surface of the following form: 
\begin{equation}
\label{eq:6}
\boldsymbol r^TQ\boldsymbol r+G=0
\end{equation}
where
$$Q=L\boldsymbol u\boldsymbol u^TL-(\boldsymbol u^TL\boldsymbol u)L$$
$$G=\boldsymbol u^TL\boldsymbol u$$
in which $L=\textrm{Diag}\left\{[1/a^2\quad 1/b^2\quad 1/c^2]^T\right\}$. Parameters $a$, $b$, and $c$ are the lengths of the ellipsoid semi-principal axes. The derivation of Eqn. \eqref{eq:6} is provided in \ref{S:7}. However, it can be intuitively shown that the locus of the possible position vectors is a cylinder (Fig. \ref{fig:2}).

Consider a horizon vector $\boldsymbol u^{RSW}$ defined in the RSW coordinate system. The RSW coordinate is defined such that its $x$ axis is in the direction of the position vector $\boldsymbol r$, the $z$ axis towards the orbital angular momentum vector of the satellite, and the $y$ axis completes the right-handed coordinate system. The direction of the horizon vector is measured by the horizon sensor and the selected vector is not necessarily a unit vector. Introducing $\boldsymbol u^{RSW}$ by spherical angles yields:
\begin{equation}
\label{eq:7}
\boldsymbol u^{RSW}=\|\boldsymbol u^{RSW}\|
\left (
  \begin{tabular}{c}
  $\sin{\theta^{RSW}}$ \\
  $\cos{\theta^{RSW}}$ $\sin{\phi^{RSW}}$ \\
  $\cos{\theta^{RSW}}$ $\cos{\phi^{RSW}}$
  \end{tabular}
\right )
\end{equation}
where, $\theta^{RSW}$ and $\phi^{RSW}$ are defined with respect to the axes of the RSW coordinate system. The angle $\phi^{RSW}$ is assumed to be predefined for the satellite. In Eqn. \eqref{eq:7}, since the value of $\|\boldsymbol u^{RSW}\|$ is not assigned, it is assumed to be $\|\boldsymbol u^{RSW}\|=\sec{\theta^{RSW}}$ and then
\begin{equation}
\label{eq:8}
u_R=\tan{\theta^{RSW}}\quad u_S=\sin{\phi^{RSW}}\quad u_W=\cos{\phi^{RSW}}
\end{equation}
in which $\boldsymbol u^{RSW}=[u_R\quad u_S\quad u_W]^T$. The vector $\boldsymbol u^{CBCS}=[u_x^{CBCS}$ $u_y^{CBCS}$ $u_z^{CBCS}]^T$ defined in CBCS can be related to $\boldsymbol u^{RSW}$ as follows: 
\begin{equation}
\label{eq:9}
\boldsymbol u^{CBCS}=C^{CBCS}_{RSW}\boldsymbol u^{RSW}
\end{equation}
in which $C^{CBCS}_{RSW}$ is the rotation matrix from RSW to CBCS and can be shown in the following form: 
\begin{equation}
\label{eq:10}
C^{CBCS}_{RSW}=
\left (
  \begin{tabular}{ccc}
  $C_{xR}$ & $C_{xS}$ & $C_{xW}$ \\
  $C_{yR}$ & $C_{yS}$ & $C_{yW}$ \\
  $C_{zR}$ & $C_{zS}$ & $C_{zW}$ \\
  \end{tabular}
\right )
\end{equation}
Substituting Eqn. \eqref{eq:9} into Eqn. \eqref{eq:6}, using Eqns. \eqref{eq:8} and \eqref{eq:10}, $\theta^{RSW}$ can be found as follows:
\begin{equation}
\label{eq:11}
\theta^{RSW}=\textrm{tan}^{-1}(\frac{-B\pm \sqrt{B^2-4AC}}{2A})
\end{equation}
where,
$$A=\boldsymbol r^TL\boldsymbol c_1\boldsymbol c_1^TL\boldsymbol r+\boldsymbol c_1^TL\boldsymbol c_1(1-\boldsymbol r^TL\boldsymbol r)$$
$$B=\boldsymbol r^TL(\boldsymbol c_1\boldsymbol c_2^T+\boldsymbol c_2\boldsymbol c_1^T)L\boldsymbol r+2\boldsymbol c_1^TL\boldsymbol c_2(1-\boldsymbol r^TL\boldsymbol r)$$
$$C=\boldsymbol r^TL\boldsymbol c_2\boldsymbol c_2^TL\boldsymbol r+\boldsymbol c_2^TL\boldsymbol c_2(1-\boldsymbol r^TL\boldsymbol r)$$
$$\boldsymbol c_1=
\left (
  \begin{tabular}{ccc}
  $C_{xR}$ \\
  $C_{yR}$ \\
  $C_{zR}$ \\
  \end{tabular}
\right )
$$
$$\boldsymbol c_2=
\left (
  \begin{tabular}{cc}
  $C_{xS}$ & $C_{xW}$ \\
  $C_{yS}$ & $C_{yW}$ \\
  $C_{zS}$ & $C_{zW}$
  \end{tabular}
\right )
\left (
  \begin{tabular}{c}
  $\sin{\phi^{RSW}}$ \\
  $\cos{\phi^{RSW}}$
  \end{tabular}
\right )
$$

Finally, the horizon unit vector is defined as:
\begin{equation}
\label{eq:12}
\boldsymbol u=\frac{\boldsymbol u^{CBCS}}{\|\boldsymbol u^{CBCS}\|} \equiv \cos{\theta^{RSW}}\boldsymbol u^{CBCS}
\end{equation}
and,
\begin{equation}
\label{eq:13}
\phi=\textrm{tan}^{-1}(\frac{u_y}{u_x}) \quad \theta={\sin}^{-1}(u_z)
\end{equation}

The summarized algorithm for obtaining the output of $\boldsymbol h(\boldsymbol x)$ in the simulation is tabulated in Table \ref{table:1}. 

\begin{table}[h]
\caption{Algorithm of the measurement model}
\centering
\begin{tabular}{l}
\hline
\textbf{Input}:  State vector $\boldsymbol x$, and angle $\phi^{RSW}$ \\
\textbf{Output}: Observed vector $\boldsymbol z=\boldsymbol h(\boldsymbol x)$ \\
\textbf{1}- Find $C_{RSW}^{CBCS}$ using state vector $\boldsymbol x$; \\
\textbf{2}- Calculate $\theta^{RSW}$ from Eqn. \eqref{eq:11};  \\
\textbf{3}- Calculate the unit vector $\boldsymbol u$ from Eqn. \eqref{eq:12}; \\
\textbf{4}- Find the nominal measurements, $\boldsymbol h(\boldsymbol x)$, from Eqn. \eqref{eq:13}; \\
\textbf{Return} $\boldsymbol z=\boldsymbol h(\boldsymbol x)$\\
\hline
\end{tabular}
\label{table:1}
\end{table}

It should be noted that in an applicable scenario, the horizon sensor finds the horizon vectors by searching the space for CB horizons. Thus, the angle $\phi^{RSW}$ is variable based on the horizon scan direction rotation and the semi-angle of its scanning cone. However, without loss of generality, it is assumed in this study that the angle $\phi^{RSW}$ is predefined. This assumption can make the simulations more efficient in terms of computational effort.

\section{Unscented Kalman Filtering}
\label{S:4}

Suppose the discretized process and measurement models be stated as follows:
\begin{equation}
\label{eq:14}
\boldsymbol x_{k+1}=\boldsymbol f_d(\boldsymbol x_k)+\boldsymbol w_k
\end{equation}
\begin{equation}
\label{eq:15}
\boldsymbol z_k=\boldsymbol h_d(\boldsymbol x_k)+\boldsymbol v_k
\end{equation}
where $\boldsymbol w_k\sim \mathcal{N}([0]_{12\times1},Q_k)$, $\boldsymbol v_k\sim \mathcal{N}([0]_{2\times1},R_k)$, and $\boldsymbol f_d(\boldsymbol x_k)$ is the discretized form of $\boldsymbol f(\boldsymbol x)$ that is calculated using the 4th order Runge-Kutta method with a specified time step. Function $\boldsymbol h_d(\boldsymbol x_k)$ can be extracted directly from the measurement algorithm of Table \ref{table:1} by substituting $\boldsymbol x=\boldsymbol x_k$. The UKF algorithm initializes with a $n$-dimensional state vector $\hat{\boldsymbol x}^{+}_0=\mathcal{E}(\boldsymbol x_0)$ as the initial estimate and an $n\times n$ matrix $P^{+}_0=\mathcal{E}[(\boldsymbol x_0-\hat{\boldsymbol x}^{+}_0)(\boldsymbol x_0-\hat{\boldsymbol x}^{+}_0)^T]$ as the initial covariance matrix. For $i=1,...,2n$ sigma points are calculated as \begin{equation}
\label{eq:16}
\hat{\boldsymbol x}^{(i)}_{k-1}=\hat{\boldsymbol x}^{+}_{k-1}+\tilde{\boldsymbol x}^{(i)}
\end{equation}
such that for $i=1,...,n$
\begin{equation}
\label{eq:17}
\tilde{\boldsymbol x}^{(i)}=\left(\sqrt{nP^+_{k-1}}\right)^T_i
\end{equation}
and for $i=n+1,...,2n$
\begin{equation}
\label{eq:18}
\tilde{\boldsymbol x}^{(i)}=-\left(\sqrt{nP^+_{k-1}}\right)^T_i
\end{equation}
in which $\sqrt{nP^+_{k-1}}$ is the matrix square root of $nP^+_{k-1}$ (that is calculated by Cholesky factorization) and $\left(\sqrt{nP^+_{k-1}}\right)_i$ is the ith row of $\sqrt{nP^+_{k-1}}$. Using the noise-less form of Eqn. \eqref{eq:14}, sigma points at the kth step are evaluated as $\hat{\boldsymbol x}^{(i)}_{k}=\boldsymbol f_d(\hat{\boldsymbol x}^{(i)}_{k-1})$. Combining the vectors $\hat{\boldsymbol x}^{(i)}_{k}$ the \textit{a priori} state estimate and the corresponding covariance matrix at the kth step are obtainable as
\begin{equation}
\label{eq:19}
\hat{\boldsymbol x}^{-}_k=\frac{1}{2n}\sum_{i=1}^{2n}\hat{\boldsymbol x}^{(i)}_k
\end{equation}
\begin{equation}
\label{eq:20}
P^{-}_k=\frac{1}{2n}\sum_{i=1}^{2n}(\hat{\boldsymbol x}^{(i)}_k-\hat{\boldsymbol x}^{-}_k)(\hat{\boldsymbol x}^{(i)}_k-\hat{\boldsymbol x}^{-}_k)^T+Q_{k-1}
\end{equation}

Now, for a measurement update, the sigma points are calculated similar to Eqns. \eqref{eq:16} to \eqref{eq:18} but with the substitution of the symbols ``$k-1$'' and ``$+$'' with ``$k$'' and ``$-$'', respectively. The sigma points can be transferred using the noise-less form of the measurement model in Eqn. \eqref{eq:15} as $\hat{\boldsymbol z}^{(i)}_k=\boldsymbol h_d(\hat{\boldsymbol x}^{(i)}_k)$. So, the predicted measurement and the associated covariance at the $k$th step is
\begin{equation}
\label{eq:21}
\hat{\boldsymbol z}_k=\frac{1}{2n}\sum_{i=1}^{2n}\hat{\boldsymbol z}^{(i)}_k
\end{equation}
\begin{equation}
\label{eq:22}
P_z=\frac{1}{2n}\sum_{i=1}^{2n}(\hat{\boldsymbol z}^{(i)}_k-\hat{\boldsymbol z}_k)(\hat{\boldsymbol z}^{(i)}_k-\hat{\boldsymbol z}_k)^T+R_{k}
\end{equation}

The cross covariance between $\hat{\boldsymbol x}^{-}_k$ and $\hat{\boldsymbol z}_k$ is obtained as follows:
\begin{equation}
\label{eq:23}
P_{xz}=\frac{1}{2n}\sum_{i=1}^{2n}(\hat{\boldsymbol x}^{(i)}_k-\hat{\boldsymbol x}^{-}_k)(\hat{\boldsymbol z}^{(i)}_k-\hat{\boldsymbol z}_k)^T
\end{equation}

Finally, the measurement update of the state (and parameter) estimation can be obtained as the following procedure:
\begin{equation}
\label{eq:24}
K_k=P_{xz}P_z^{-1}
\end{equation}
\begin{equation}
\label{eq:25}
\hat{\boldsymbol x}_k^{+}=\hat{\boldsymbol x}_k^{-}+K_k(\boldsymbol z_k-\hat{\boldsymbol z}_k)
\end{equation}
\begin{equation}
\label{eq:26}
P^+_k=P^-_k+K_kP_zK_k^T
\end{equation}

\section{Simulation Results}
\label{S:5}

\begin{figure}[H]
\centering\includegraphics[width=0.6\linewidth]{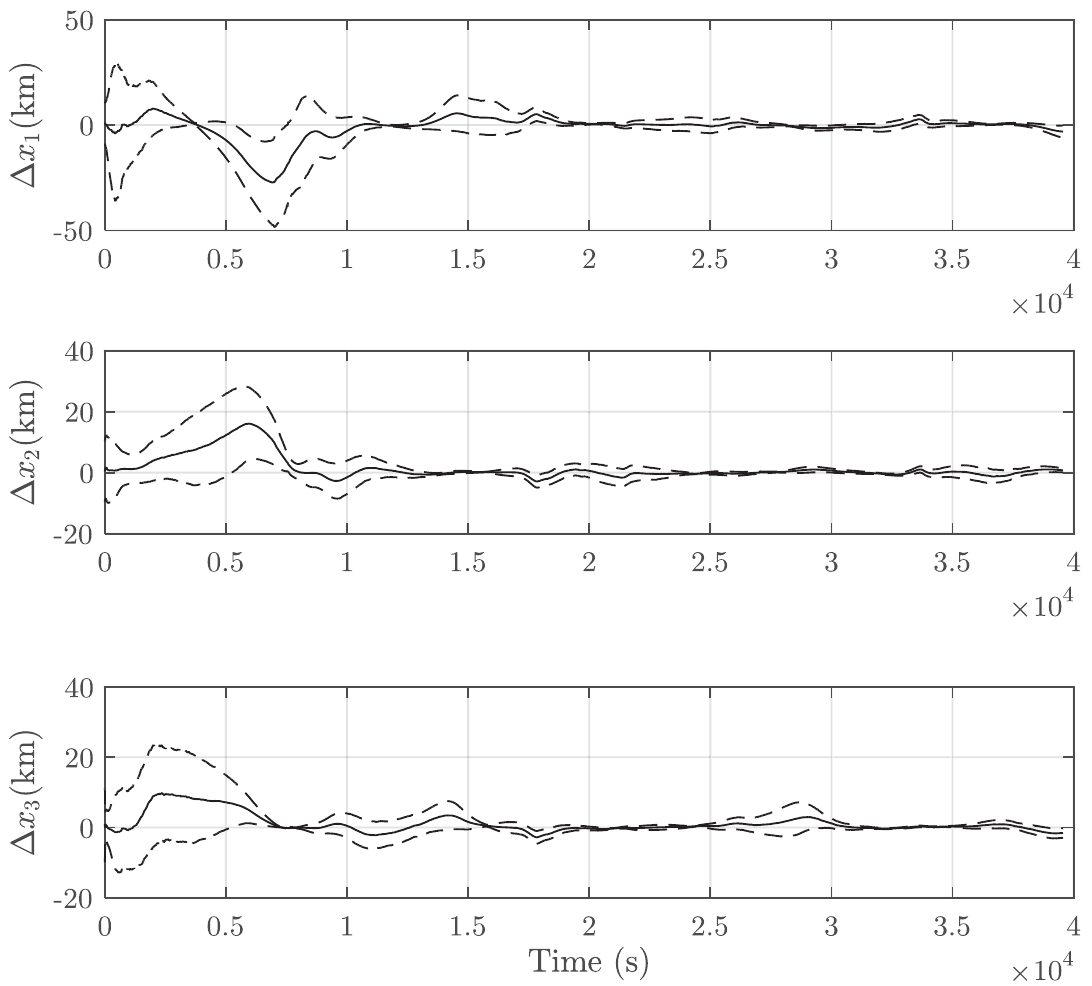}
\caption{Estimation error of spacecraft position vector ($\Delta \boldsymbol x_{1:3}$); mean value (solid); mean value$\pm$standard deviation (dashed).}
\label{fig:3}
\end{figure}

\begin{figure}[H]
\centering\includegraphics[width=0.6\linewidth]{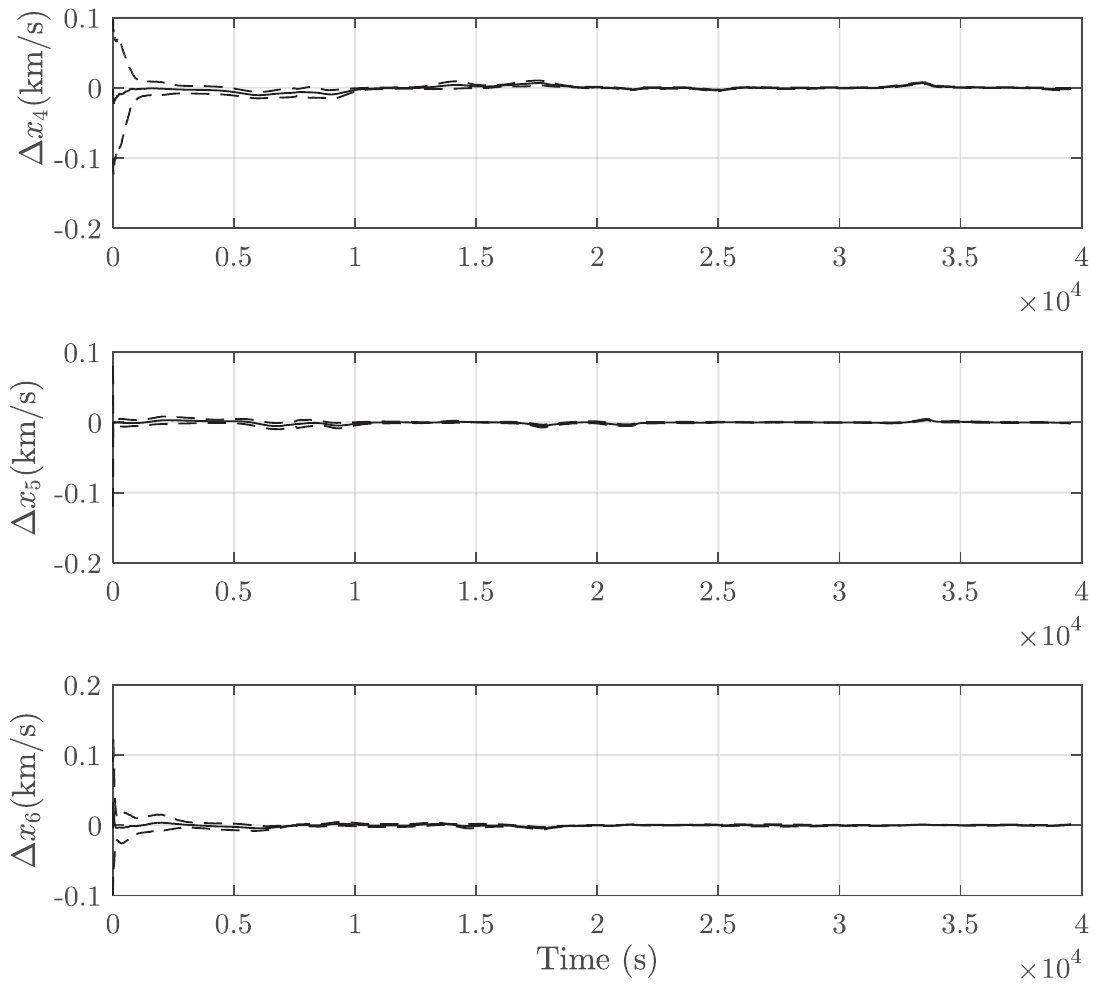}
\caption{Estimation error of spacecraft velocity vector ($\Delta \boldsymbol x_{4:6}$); mean value (solid); mean value$\pm$standard deviation (dashed).}
\label{fig:4}
\end{figure}

\begin{figure}[H]
\centering\includegraphics[width=0.6\linewidth]{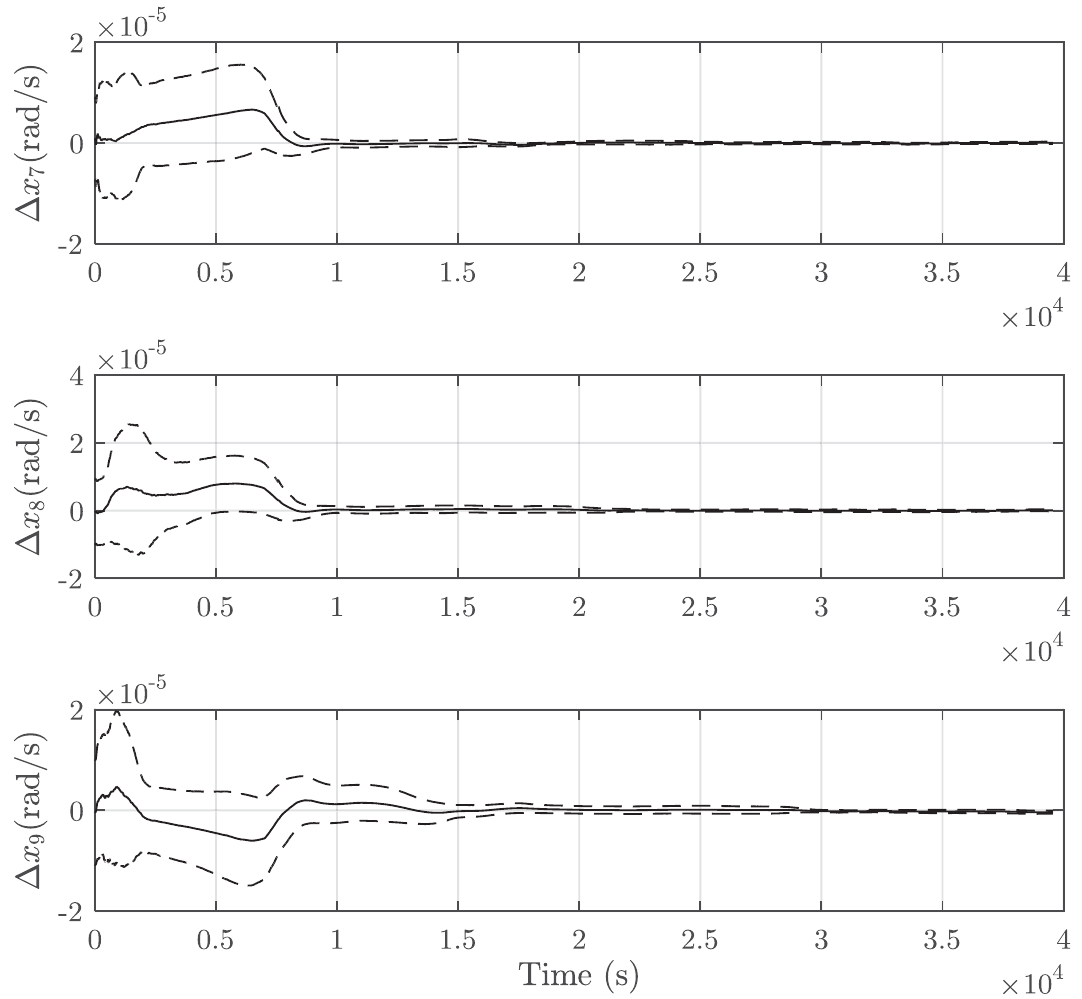}
\caption{Estimation error of Ceres rotation vector ($\Delta \boldsymbol x_{7:9}$); mean value (solid); mean value$\pm$standard deviation (dashed).}
\label{fig:5}
\end{figure}

\begin{figure}[H]
\centering\includegraphics[width=0.6\linewidth]{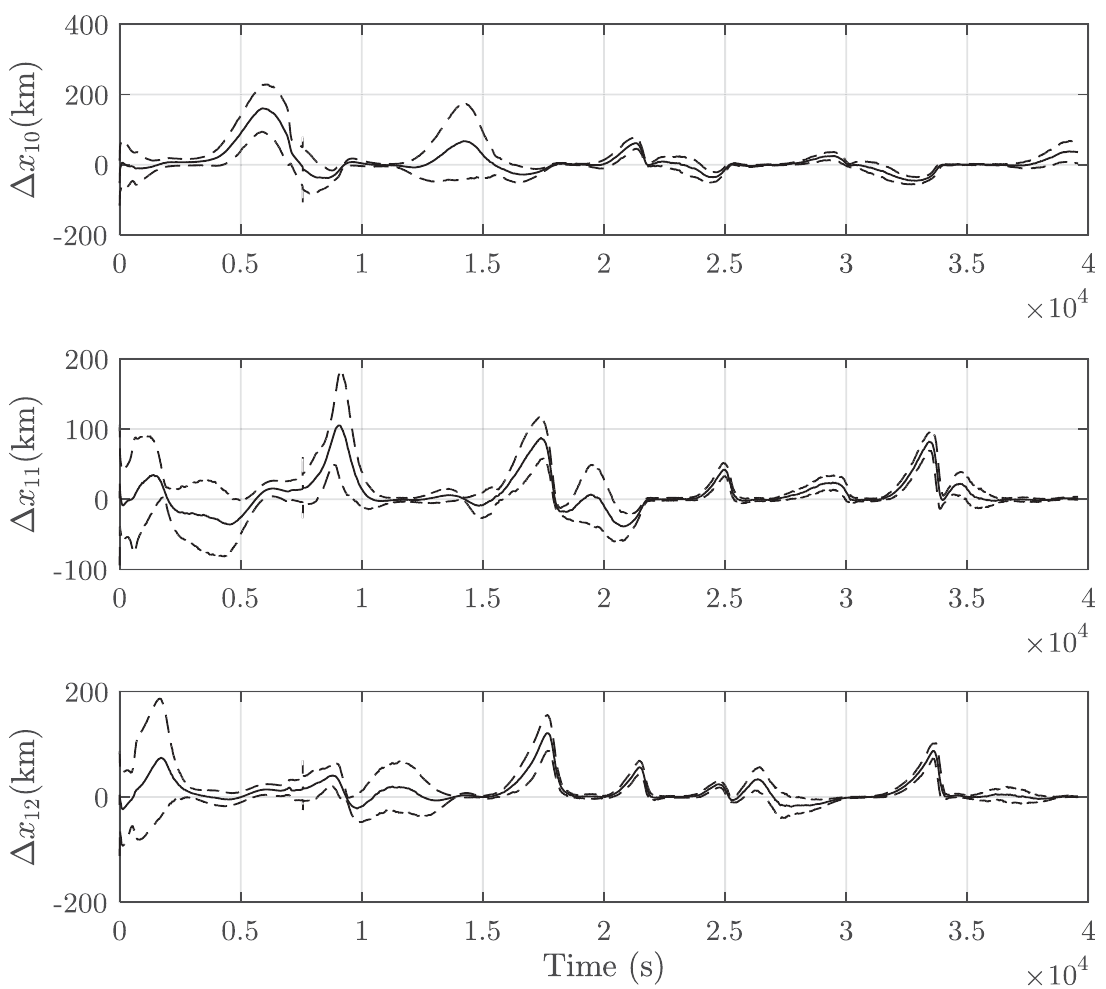}
\caption{Estimation error of Ceres semi-principal axis lengths ($\Delta \boldsymbol x_{10:12}$); mean value (solid); mean value$\pm$standard deviation (dashed).}
\label{fig:6}
\end{figure}

With the process and measurement models that have been introduced in the previous section, the standard UKF has been utilized to estimate the state of the satellite as well as the angular velocity of the celestial body, and its semi-principal axes lengths. Thus, vector $\boldsymbol x$ is including the position vector, $\boldsymbol x_{1:3}\equiv \boldsymbol r$, the velocity vector, $\boldsymbol x_{4:6}\equiv \dot{\boldsymbol r}$, the angular velocity of the celestial body, $\boldsymbol x_{7:9}$, and the semi-principal axes lengths of the celestial body, $\boldsymbol x_{10:12}$. For the case of this study, the celestial body is assumed to be the asteroid Ceres. The specified time step for Runge-Kutta numerical integration is considered to be $\boldsymbol \tau_{RK4}=0.2$ s. The moment of inertia $J$ is calculated as follows assuming a uniform mass distribution:
\begin{equation}
\label{eq:27}
J=\frac{1}{5}M_{Ceres}\textrm{Diag}\left\{[b^2+c^2 \quad c^2+a^2 \quad a^2+b^2]^T\right\}
\end{equation}
in which $M_{Ceres}=9.393 \times 10^{20}$ kg is the mass of Ceres. The true semi-principal axes lengths of Ceres are $a=482.6$ km, $b=480.6$ km, and $c=445.6$ km. In the estimation procedure, $\hat{J}=\mathcal{E}(J)$ is obtained from estimated semi-principal axes lengths, $\hat{a}$, $\hat{b}$, and $\hat{c}$. The true angular velocity of Ceres is $\boldsymbol \omega=[0 \quad 0 \quad 1.9234\times 10^{-4}]^T$ rad/s. The scenario of the spacecraft orbital motion around Ceres is defined to be a circular polar orbit with a semi-major axis of $857.6$ km. The initial estimations of state, and parameters, are generated by adding a random normally distributed vector to the true states, and parameters in each Monte-Carlo simulation, $\hat{\boldsymbol x}_0=\boldsymbol x_0+\Delta \boldsymbol x_0$, $\Delta \boldsymbol x_0 \sim \mathcal{N}([0]_{12\times1},P_0)$. The initial estimation covariance is
$$P_0=\textrm{Diag}\left\{\left[10^2[1]_{1\times3} \quad 10^{-2}[1]_{1\times3} \quad 10^{-10}[1]_{1\times3}\quad 10^4[1]_{1\times3}\right]^T\right\}$$
The covariance of process noise is defined as
$$Q_k=\textrm{Diag}\left\{\left[[0]_{1\times3} \quad 10^{-16}[1]_{1\times3} \quad [0]_{1\times3}\quad [0]_{1\times3}\right]^T\right\}$$
that applies an acceleration uncertainty with standard deviation of $10^{-8}$ km/s$^2$ to the orbital motion of the spacecraft around the asteroid. The covariance of measurement noise is $R_k=3.046\times 10^{-6} \mathbb{I}_2$ that corresponds to a standard deviation of $0.1^{\circ}$ for the measured angles of the horizon detector. The angle $\phi^{RSW}$ chooses the values of $0^{\circ}$, and $30^{\circ}$ periodically, with a $1$ s gap. It means that the scan direction rotates at a speed of $0.5$ revolution-per-second and in this way non-simultaneous measurements of the horizon sensor can be used. By predetermining $\phi^{RSW}$, there is no need for simulation of the spacecraft body attitude. The estimation errors, $\Delta \boldsymbol x=\hat{\boldsymbol x}-\boldsymbol x$, are shown in Fig. \ref{fig:3} for the position, and in Fig. \ref{fig:4} for the velocity of the spacecraft. The mean estimation errors as well as their standard deviations are calculated using Monte-Carlo simulations. It can be seen that although the initial position and velocity estimation errors are large, the estimation errors converge to near zero values. The estimation errors for the angular velocities and the semi-principal axes lengths of Ceres are shown in Figs. \ref{fig:5} and \ref{fig:6}, respectively. It is clear that the estimation of the asteroid angular velocity is very accurate and rapidly converges.

\begin{figure}[H]
\centering\includegraphics[width=0.8\linewidth]{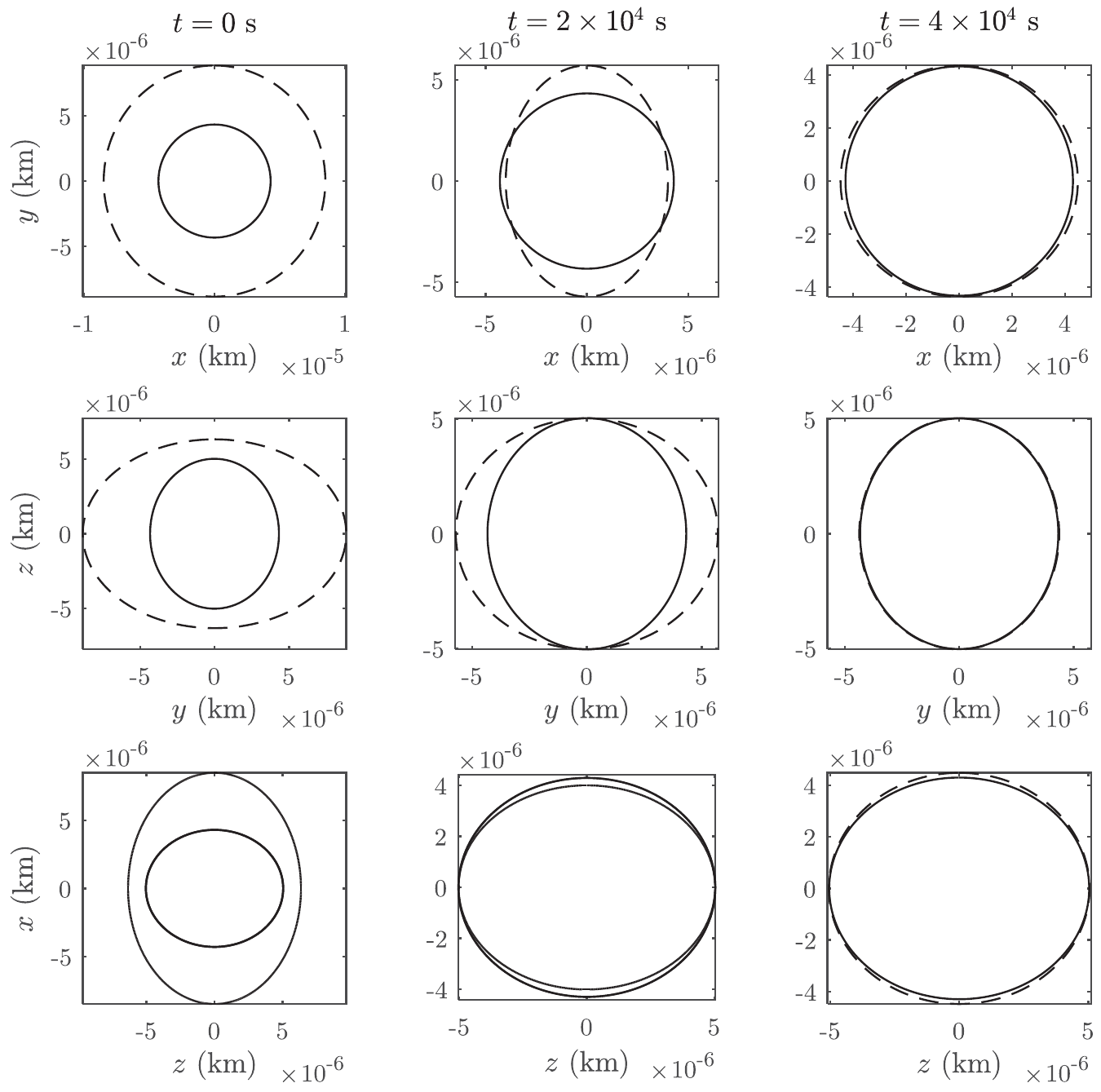}
\caption{Estimated (dashed) against real (solid) geometry of Ceres at three time instances of $t=0$, $2\times10^4$, and $4\times10^4\text{ s}$.}
\label{fig:7}
\end{figure}

The evolution of the estimated geometry of Ceres is shown in Fig. \ref{fig:7} for the $x-y$, $y-z$, and $z-x$ planes, respectively for a typical simulation. The geometry is drawn using the estimated semi-principal axis of Ceres at $t=0$ s (the initial estimation), $t=2\times 10^4$ s, and $t=4\times 10^4$ s. Fig. \ref{fig:7} shows the convergence of the estimated CB geometry to the real geometry of Ceres (shown in black) under the UKF by the use of the proposed algorithm. 

\section{Conclusion}
\label{S:6}

A process and measurement model was proposed for an orbit and parameter estimation of a spacecraft around a celestial body using non-simultaneous horizon detector measurements. The celestial body rotational velocity and its semi-principal axes lengths were estimated along with the satellite orbital position and velocity. The proposed method demonstrated that using UKF the estimated state and parameters converge to the real values, even though the geometry and the rotational velocity of the celestial body are not initially known. The process model contained the gravitational perturbations as a function of celestial body geometry. The measurement model provided the opportunity to consider the time gaps between non-simultaneous measurements. These results show the possibility of using a horizon detector for determining some geometric parameters of the celestial bodies by assuming a homogenous tri-axial ellipsoid model of the celestial body. However, in future studies, the mass and the mass distribution of the celestial body can also be taken as unknown parameters and can possibly be estimated using horizon sensor measurements and the nonlinear model of the system.

\appendix
\section{Derivation of Relation Between $\boldsymbol u$ and $\boldsymbol r$}
\label{S:7}

First, suppose that the center of the ellipsoid is located in the center of the coordinate system, and its semi-principal axes are the same as the axes directions of the coordinate system. The parametric equation of a line in the direction of $\boldsymbol u$ that passes through $\boldsymbol r$ can be written as
\begin{equation}
\label{a:1}
\boldsymbol \tau= \rho \boldsymbol u +\boldsymbol r
\end{equation}

The vector $\boldsymbol \tau$ points to the position of any point along the line $\boldsymbol u$ with the distance $\rho$ from the spacecraft. The intersection of the line, defined by Eqn. \eqref{a:1}, and the ellipsoid occurs at $\boldsymbol \tau=\boldsymbol \tau_0$ that can be obtained by substituting Eqn. \eqref{a:1} in the ellipsoid equation (Fig. \ref{fig:2}):
\begin{equation}
\label{a:2}
\rho \boldsymbol u^TL\boldsymbol u+2 \rho \boldsymbol u^TL\boldsymbol r+\boldsymbol r^TL\boldsymbol r=1
\end{equation}
Eqn. \eqref{a:2} is a quadratic equation in terms of $\rho$. In order for the line of Eqn. \eqref{a:1} to be tangent to the ellipsoid, the following equation should be satisfied:
\begin{equation}
\label{a:3}
(\boldsymbol u^TL\boldsymbol r)^2-(\boldsymbol u^TL\boldsymbol u)(\boldsymbol r^TL\boldsymbol r-1)=0
\end{equation}
Eqn. \eqref{a:3} can be simplified to the form presented as Eqn. \eqref{eq:6}.

\bibliography{mybibfile}

\end{document}